\documentclass[fleqn,usenatbib]{mnras}         
\usepackage{newtxtext,newtxmath}
\usepackage[T1]{fontenc}
\usepackage{graphicx}   
\usepackage{natbib}   
       
\usepackage{amssymb}
\usepackage{amsmath}
\usepackage{ulem}
\title[\(\eta\) does not affect mass measurement]{Kinematical small-scale fluctuations do not affect the measurement of the dynamical mass of galaxies}

\author[Z. Zhong \& G. Zhao]{
Zehao Zhong,$^{1,2}$ and Gang Zhao$^{1,2}$\thanks{E-mail: gzhao@nao.cas.cn}
\\
$^{1}$CAS Key Laboratory of Optical Astronomy, National Astronomical Observatories, Chinese Academy of Sciences, Beijing, 100101, China\\
$^{2}$School of Astronomy and Space Science, University of Chinese Academy of Sciences, Beijing, 100049, China
}

\date{Accepted 2024 February 16. Received 2024 February 08; in original form 2023 June 29}

\pubyear{2023}

\begin{document}
\label{firstpage}
\pagerange{\pageref{firstpage}--\pageref{lastpage}}
\maketitle

\begin{abstract}
The stellar kinematics of low-mass galaxies are usually observed to be very unsmooth with significant kinematical fluctuations in small scales, which cannot be consistent with the projected centrosymmetric stellar kinematics obtained from commonly used dynamical models. In this work, we aim to test whether the high degree of kinematical fluctuations affects the dynamical mass estimate of galaxies. We use the asymmetry parameter \(\eta\) obtained from the \(180^{\circ}\) rotation self-subtraction of stellar kinematics of galaxies to quantify the degree of kinematical small-scale fluctuations. We use TNG50 numerical simulation to construct a large sample of mock galaxies with known total masses, and then obtained the virial dynamical mass estimator of these mock galaxies. { We find that the dynamical masses within three-dimensional \(R_{\rm e}\) to the mock galaxy centres are overall averagely accurate within around 0.1 dex under the symmetric assumption, while \(R_{\rm e}\) means the projected circularized half-stellar mass radius in this work. We study the local virial mass estimation bias for mock galaxies of different \(\eta\). The maximum bias difference of two \(\eta\) bins is around 0.16 dex, which with other local biases may help apply the observational virial mass estimators obtained from massive galaxies to other types of galaxies.} We find that the Spearman's \(\rho\) of \(\eta\) with the {intrinsic mass estimation deviations is near zero if the local bias is eliminated properly}. The results indicate that even for low-mass galaxies, the existence of high degree of kinematical small-scale fluctuations does not affect the measurement of the dynamical mass of galaxies.
\end{abstract}	

\section{Introduction}
Mass is one of the most fundamental and important parameters involved in almost every aspect of studying galaxies. {Stellar mass is generally obtained through the stellar population synthesis model or related empirical estimation functions \citep{2001ApJ...550..212B,2003ApJS..149..289B,2011MNRAS.418.1587T}. However, most galaxies generally contain large amounts of dark matter, so the total mass is usually measured by dynamical methods and is therefore often called dynamical mass.} The precise dynamical mass not only plays a very important role in the study of the formation and evolution of galaxies, but can also be used to study the matter distribution and dark matter content of galaxies \citep{2006MNRAS.366.1126C,2013MNRAS.432.1709C}. Different mass distributions of galaxies indicate different gravitational fields, which result in different kinematics. {{People usually use stellar kinematics to construct dynamical models.}} The simplest method is the {virial} theorem \citep{2008gady.book.....B}.  \citet{2006MNRAS.366.1126C} used integral field {spectroscopy} (IFS) to observe the kinematics of massive galaxies, and build different dynamical models to measure their galaxy dynamical masses. {Their results demonstrated} that the corresponding virial mass measurement results are relatively reliable and unbiased. {\citet{2022ApJ...936....9V} also investigated the virial mass estimate. Although they have found a small systematic offset of their mass scale with that from \(\rm ATLAS^{3D}\) \citep{2013MNRAS.432.1709C}, they conclude that their virial mass estimate is effective with a systematic uncertainty of at most 0.1 dex.}

Due to the {complex nature} of galaxies, the simple virial {theorem} cannot {{fully capture} the stellar kinematics} of actual galaxies, so more complex models are necessary. These dynamical models are generally spherical model, axisymmetric model or triaxial model. Although the virial model can also be regarded as a spherical model, the spherical model is far more than that, and it is the earliest developed galaxy dynamical model that includes many kinds. Nowadays, the {{spherical S\'ersic profiles \citep{1963BAAA....6...41S} are one of the most widely used profiles to describe the optical distribution of galaxies, which can assist in dynamical calculations of spherical models.}} For the axisymmetric model, Jeans Anisotropic Modelling (JAM) is well known \citep{2008MNRAS.390...71C,2020MNRAS.494.4819C}. JAM is widely used in various galaxy IFS surveys \citep{2013MNRAS.432.1709C,2017ApJ...838...77L,2019MNRAS.490.2124L,2023MNRAS.522.6326Z} and also {in studies of individual galaxies} \citep{2013MNRAS.436.2598W,2016MNRAS.463.1117Z}. For {triaxial models}, Schwarzschild constructed a numerical stellar system in dynamical equilibrium model and studied the corresponding stellar kinematics in detail \citep{1979ApJ...232..236S}. This is one of the earliest and is now the most widely used triaxial system model. In addition, since {triaxial Schwarzschild models have more degeneracies, axisymmetric models are more popular among many studies and have been shown to infer accurate dynamical masses \citep{2012MNRAS.424.1495L,2013MNRAS.432.1709C,2018MNRAS.477..254L}. All these models assume a certain degree of symmetry. After projection on to a two-dimensional plane, the second moment of the velocity distribution obtained from stellar kinematics is required to be centrosymmetric. Most massive early-type galaxies (ETGs) are observed as well centrosymmetric kinematical patterns, providing good evidence for this \citep{2013MNRAS.432.1709C}.}

{Since dynamical mass cannot be measured directly,} we can never know the true mass of the galaxy, so we {cannot} compare it with model mass to obtain the measurement accuracy. Fortunately, the {projected velocity field of a galaxy} is a relatively direct measurement value, and people generally compare the observed velocity field with that predicted by the model. If the two are in good agreement, then we can believe that the model mass is credible. This is very successful when measuring the dynamical mass of massive galaxies, their stellar velocity fields are all centrosymmetric or near-centrosymmetric images, and the observation results are in good agreement with the model predictions \citep{2006MNRAS.366.1126C,2013MNRAS.432.1709C}.

However, the stellar kinematics of galaxies are not all regular. The morphological classification of stellar kinematics can be divided into {different} classes \citep{2011MNRAS.414.2923K,2016ARA&A..54..597C}, among which there are galaxies with {non-regular} rotation. {The galaxy classification of non-regular rotators is also related to high-order stellar kinematics \citep{2017ApJ...835..104V} and relevant to galaxy merger histories according to simulations \citep{2014MNRAS.444.3357N}.} The {stellar kinematics} of {non-regular rotation or non-rotation} galaxies {are generally not consistent} with regular and smooth model predictions. In addition, things {are different} for low-mass galaxies, {which in this paper refer to galaxies of \(\log(M/M_{\rm \odot})\leq 9.5\), approximately the biggest stellar mass of dwarf galaxies}. They usually contain lots of small patches in their velocity field, making them far from smoothness and kinematical symmetry. Our recent work ({{\citealt{2023ApJ...957L..12Z}, hereafter Paper I}}) has found that the unsmoothness and asymmetry of the observational stellar velocity field in low-mass galaxies are real and {cannot} be fully {explained} by measurement uncertainties. These patches are like fluctuations in the regular, smooth and symmetric pattern predicted by the ideal model, so we call them kinematical small-scale fluctuations. We further found that the fluctuation degree is closely inversely log-linearly related to the galaxy stellar surface mass density, {and the relation is among galaxies that do not show obvious optical asymmetry that usually traces environmental perturbations}. According to this relation, although the kinematics of massive galaxies {are} relatively centrosymmetric, the small-scale fluctuation degree of low-mass galaxies is {significantly higher by an order of magnitude}, causing their velocity fields to deviate far from centrosymmetric. {{Therefore, it is natural to ask whether all the centrosymmetric dynamical models introduce a bias in the dynamical mass estimate.}} In this paper, we use {the} TNG50 \citep{2019MNRAS.490.3234N,2019MNRAS.490.3196P} numerical simulation data combined with virial {theorem} to test the dynamical mass measurements. The results {show} that the presence of small-scale fluctuations does not affect dynamical mass measurements, and virial mass {measurements are} also relatively reliable for low-mass galaxies.

The paper is organized as follows: In section 2, we introduce the existence of kinematical small-scale fluctuations and the corresponding calculation methods in Paper I, as well as the simulated mock galaxies we construct from TNG50; in section 3, we briefly describe the two virial dynamical mass estimators we used; and in Section 4, we present the results of the virial mass measurement, and its independence from the small-scale fluctuations. We summarize in Section 5.

\section{Kinematical small-scale fluctuations and data}
\subsection{Kinematical small-scale fluctuations}
\begin{figure*}
	\begin{center}
	\centering	
	\includegraphics[width=\textwidth,angle=0]{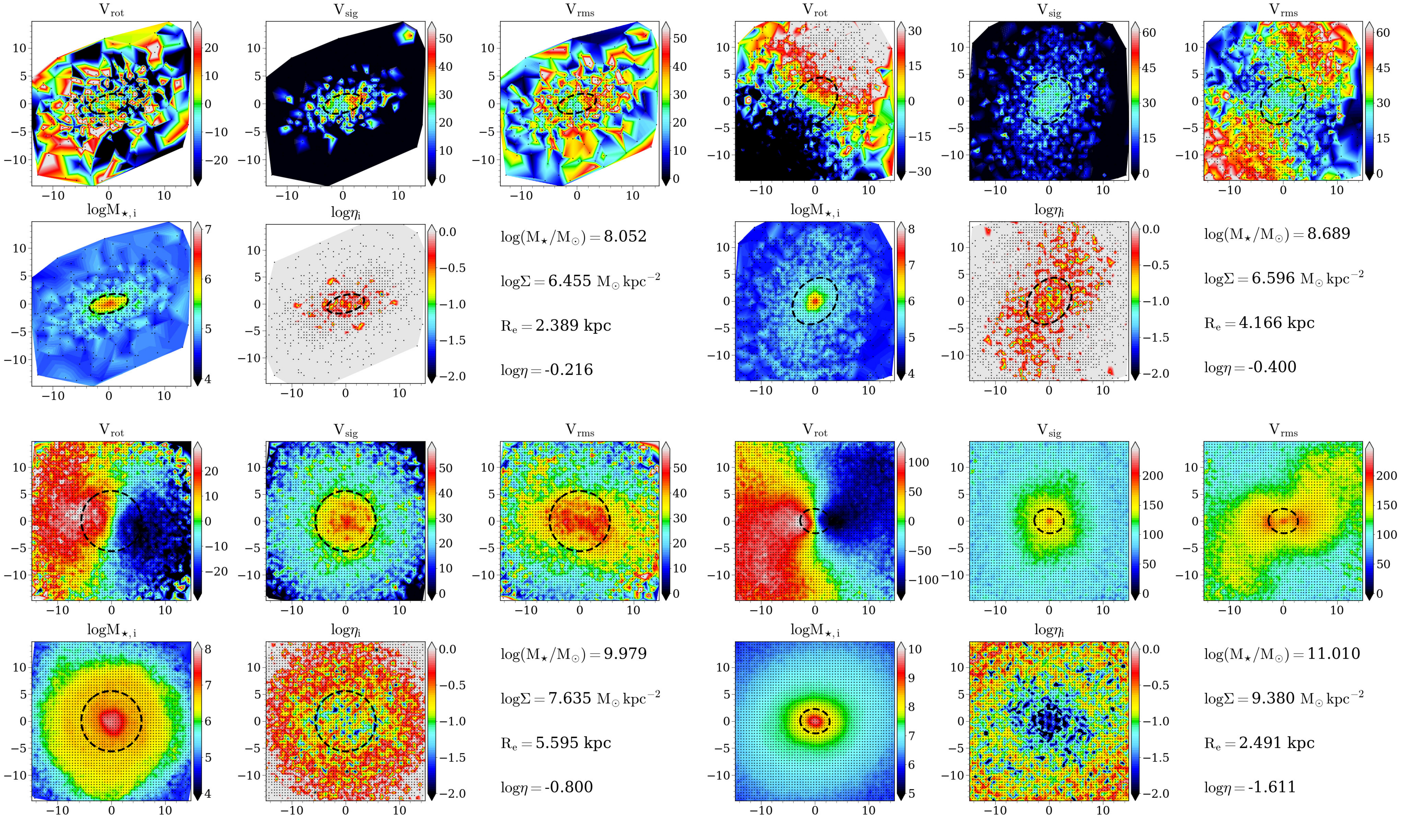}
	\end{center}
\caption{Comparison of small-scale fluctuation degrees among different TNG50 mock galaxies. The five maps of each panel represent stellar velocity, stellar velocity dispersion, total stellar velocity, {{stellar}} mass distribution, and asymmetry parameter \(\eta_{\rm i}\), respectively, of each spaxel of the mock IFU data cubes. For a better visualization, the scale of each spaxel is 0.5 kpc instead of 0.1 kpc in this figure. The parameters of each galaxy are listed in the lower right corner of each panel, which represent total stellar mass, stellar surface mass density within effective radius, effective radius, and total asymmetry parameter within effective radius, respectively. \label{fig1}}
\end{figure*}

The {symmetric} light distribution reflects the {symmetric stellar} mass distribution, {which indicates that the galaxy is in a relatively stable or self-consistent state.} {In observations,} the inner parts of galaxies, {roughly refers to within the scale of {{circularized}} effective {{(half-light)}} radius \(R_{\rm e}\),} are generally optically symmetric and relatively smooth for {different morphological-type galaxies} \citep{1995ApJ...447...82R,1998AJ....116.1163R,2000ApJ...529..886C}. Therefore, { theoretically the corresponding gravitational potential field should also be stable and have a certain degree of symmetry, and then the observed second moment velocity distribution for stellar kinematics is generally symmetric too.} So this not only is an assumption of the dynamical models, but also has some certain practical significance.
However, using different IFS surveys to observe the spatially resolved kinematics inside the galaxy, it can be found that the actual stellar velocity field is not as smooth and symmetric as given by the model. \citet{2013MNRAS.432.1709C} studied the stellar kinematics of massive {ETGs} from the \(\rm ATLAS^{3D}\) \citep{2011MNRAS.413..813C} IFS survey. {Their fig. 1} shows that even for massive galaxies, many are not so smooth in velocity fields. Fig. 1 of \citet{2019MNRAS.490.2124L} and {fig. 1} of Paper I also showed the unsmoothness and asymmetry of the galaxy kinematics in the MaNGA \citep{2015ApJ...798....7B} and SAMI \citep{2021MNRAS.505..991C} IFS surveys, respectively. In these IFS observations, the kinematical asymmetry is mainly due to the presence of small patches in the stellar velocity maps. {These patches also exist in mock galaxies. As shown in Fig. \ref{fig1}, mock galaxies with lower stellar mass or stellar surface mass density have more patches that exhibit small-scale fluctuations in stellar kinematics, and their overall fluctuation degrees are also higher.} Nevertheless, these small patches can easily be mistaken for measurement uncertainties and have not been well discussed before. In addition, galaxies in reality cannot be perfectly symmetric and must deviate somewhat. So how asymmetric is asymmetric? We also need to measure the specific amount of these patches. 

The traditional method of exploring optical symmetry is generally \(180^{\circ}\) rotation self-subtraction \citep{2000ApJ...529..886C,2003ApJS..147....1C,2004AJ....128..163L} or Fourier decomposition \citep{1995ApJ...447...82R,1998AJ....116.1163R,2010AJ....139.2097P,2017ApJS..232...21K}, and the measurement of kinematic asymmetry generally uses Fourier decomposition \citep{2006MNRAS.366..787K,2008MNRAS.390...93K,2013MNRAS.432.1768K,2017MNRAS.465..123B}. As mentioned in Paper I, Fourier decomposition requires fitting the velocity fields. For low-mass galaxies with larger measurement uncertainties, fitting results will not be so reliable. Therefore, our asymmetry parameter \(\eta\) is derived from {{the ratio of \(180^{\circ}\) rotation self-subtraction difference of the square value of root-mean-square (rms) velocity \(V_{\rm rms}\) and the original mean value}}, here \(V_{\rm rms}=\sqrt{(V_{\rm rot}^2+V_{\rm sig}^2)}\) for every velocity field spaxel (spatial pixel). {The algorithm is as \citet{zhzh2023}, and the} equation is as followed:
\begin{equation}\label{func1}
	\eta_{\rm i}=\frac{\left| V_{\rm i}^2-V_{\rm i,180}^2\right|}{(V_{\rm i}^2+V_{\rm i,180}^2)/2}
\end{equation}  
In this equation, \(\eta_{\rm i}\) is the asymmetry parameter of spaxel i, while \(V_{\rm i}\) and \(V_{\rm i,180}\) are the \(V_{\rm rms}\) of the {i\textsuperscript{th}} spaxel and its \(180^{\circ}\) rotated one. The total \(\eta\) of each galaxy is light-weighted or stellar mass-weighted average of all spaxels within galaxy effective radius \(R_{\rm e}\). Therefore \(\eta\) can characterize the asymmetric proportion of the squared total velocity in the galaxy inner part, {{which can also describe the degree of kinematical fluctuations.}} In Paper I, we also stated that the main contribution of the asymmetry parameter \(\eta\) lies in small-scale fluctuations, so \(\eta\) can be used to describe the fluctuation degree very well. We use this method to measure the asymmetry parameter \(\eta\) of simulated mock galaxies in this paper, so as to compare whether \(\eta\) affects the deviation of the virial mass from the {real} galaxy mass in TNG50.

In Paper I, we have discussed the measurement uncertainties of stellar kinematics from the SAMI survey in details {adopting the quantification of kinematical fluctuation degree.} The results show that {these small patches are just like physical fluctuations besides of measurement uncertainties}. Not only have we carried out quantitative study on the asymmetry degree, but also we obtained a very close {inversely log-linear} relationship, revealing the intrinsic correlation between the {fluctuation} degree and galaxy stellar surface mass density, {thus the \(\eta-\Sigma\) relation}. We also used both observations \citep[The SAMI survey DR3;][]{2021MNRAS.505..991C} as well as simulations \citep[TNG50;][]{2019MNRAS.490.3234N,2019MNRAS.490.3196P} to demonstrate the existence of kinematical small-scale fluctuations and also the corresponding relation. {We note that our sample galaxies from the SAMI survey in Paper I are not under galaxy merger or interaction stage, which indicates that the kinematical small-scale fluctuations are intrinsic properties within galaxies. Paper I also indicates that \(\eta\) is less related to the external environment such as galaxy volume density, but mainly related to mass distribution and dynamics within galaxies.}

{{Because \(\eta\) is calculated from the stellar kinematics of galaxies, the observational effects have large influence on the \(\eta\) as well as the \(\eta-\Sigma\) relation. In Paper I we have discussed many of them, such as measurement uncertainties, spatial resolutions of IFU spaxel, selection effects, and the point spread function (PSF). PSF has non-negligible influences on image or stellar kinematics, thus it is a common issue among many studies involving the observed stellar kinematics or dynamics of galaxies \citep{1994A&A...285..723E,2008MNRAS.390...71C,2017ApJ...835..104V,2018MNRAS.477.4711G,2020MNRAS.497.2018H}. The results of Paper I show that the \(\eta-\Sigma\) relation of observations is in good agreement with the numerical simulations after adding the influence of these various observational effects. This also indicates that the TNG50 numerical simulation can well restore the asymmetry parameter \(\eta\) of the real galaxies. Therefore, in this paper, we intend to adopt the TNG50 to investigate the intrinsic influence of \(\eta\) on dynamical mass measurement, without involving various observational effects or the additional comparison of \(\eta\) from observational and simulated galaxies.}}

\subsection{Mock galaxies from TNG50}
In order to measure the effect of asymmetry parameter \(\eta\) on the galaxy dynamical mass measurement, we need to build a model to calculate the dynamical mass of a {set} of known galaxies. {Simulation data of galaxy with known mass is very useful to assess the dynamical model and mass measurements \citep{2016MNRAS.455.3680L}. Therefore, we adopted the IllustrisTNG project \citep{2019ComAC...6....2N}, which} is currently {one of} the most advanced and widely used galaxy simulation data. It has three sets of simulations with different box sizes, among which TNG50 \citep{2019MNRAS.490.3234N,2019MNRAS.490.3196P} has the best mass resolution and is {the most suitable for studying low-mass galaxies of stellar mass \(\log(M/M_{\rm \odot})\leq 9.5\)}. Here, we use galaxies {at} zero redshift, that is, snapshot = 99 of TNG50-1. In order to reduce the influence of galaxy interactions, we only use the primary (central) {subhalo galaxy} in each group. The cosmological parameters we use are also the same as TNG, that is, the dimensionless Hubble constant \(h=0.6774\), the total matter density \(\Omega_m=0.3089\) and the dark energy density \(\Omega_{\Lambda}=0.6911\).
\begin{figure*}
	\begin{minipage}[h]{1.0\linewidth}
		\centering	
		\includegraphics[width=0.49\textwidth,angle=0]{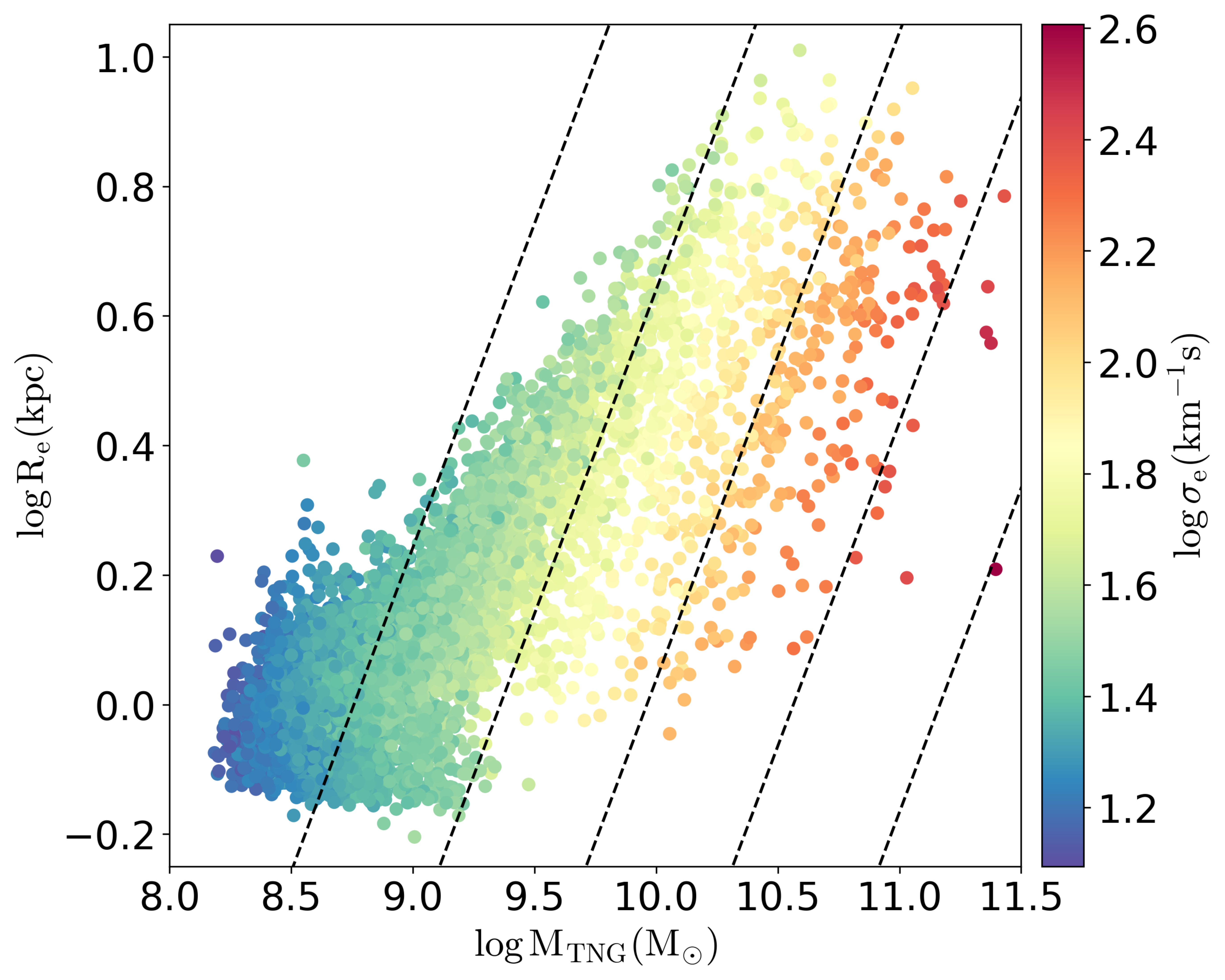}
		\includegraphics[width=0.49\textwidth,angle=0]{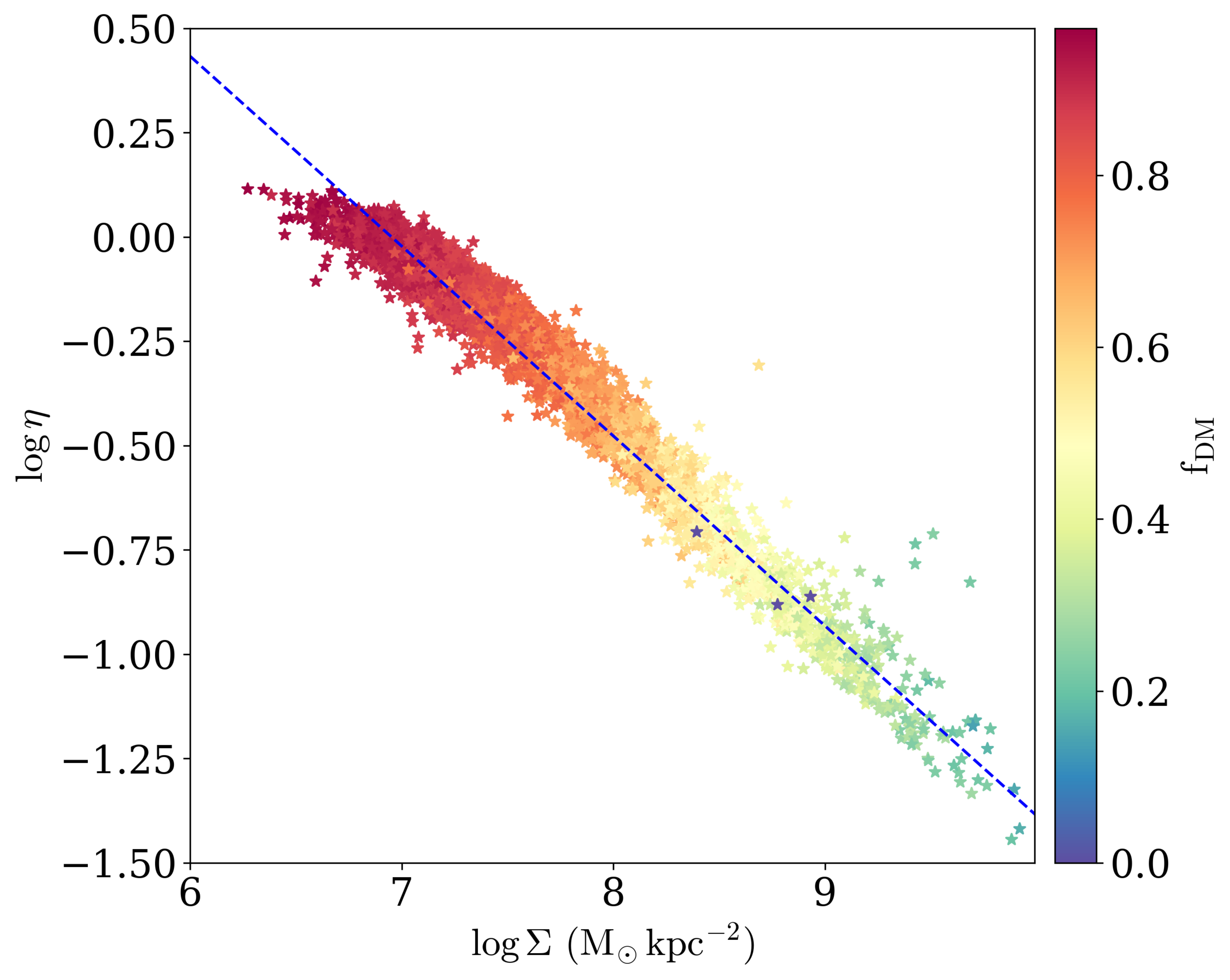}
	\end{minipage}
\caption{Mass distribution of TNG mock galaxy samples. Left panel: Mass\(-\)size plane of the samples. Here \(M_{\rm TNG}\) is the total mass of gas, stellar, and dark matter particles within three-dimensional \(R_{\rm e}\) to the mock galaxy centre. The dashed line indicates the \(\sigma_{\rm e}\) derived by the best-fitting equation (4). From left to right, the \(\sigma_{\rm e}\) are 25, 50, 100, 200, and 400 \(\rm km\,s^{-1}\) respectively. Right panel: \(\eta-\Sigma\) log-linear relation of the samples. The colourbar represents the dark matter fraction within the three-dimensional \(R_{\rm e}\). The dashed line represents the best linear fit at \(\rm scale=0.1kpc\) obtained from Paper I. The relative uncertainties of \(\eta\) are about 25\% by bootstrapping. \label{fig2}}
\end{figure*}

Each galaxy in TNG50 is composed of a {set} of particles. In order to simulate the observed two-dimensional image, we need to integrate the particles on to the two-dimensional grid. {We do not involve particle luminosity here, but perform a simpler way that use the integrated stellar mass cube as the flux cube. It is equivalent to the stellar mass-to-light ratio being 1, thus {we ignore the difference in stellar mass-to-light ratio between different galaxies, and between different regions within each galaxy. The effective radius \(R_{\rm e}\) also becomes projected circularized half-stellar mass radius in this work.} The mock stellar velocity cubes are also stellar mass-weighted in each grid. Although {there may be some systematic biases between the light-weighted and stellar mass-weighted measurement parameters \citep{2023MNRAS.518.5376D}, we considered that the difference between light-weighted and stellar mass-weighted \(\eta\) does not change the \(\eta-\Sigma\) relation significantly, while Paper I have shown that the stellar mass-weighted \(\eta-\Sigma\) relation obtained from TNG mock galaxies agree well with the light-weighted \(\eta-\Sigma\) relation obtained from observational SAMI galaxies. Therefore, adopting stellar mass-weighted does not essentially affects our goals in this paper.}} In {another aspect,} we discussed the effect of different scales (the size of grid side length) on \(\eta\) {in Paper I}. Since there exists similar \(\eta-\Sigma\) relations under different scale, and the grid scale affect less on the galaxy dynamical mass, we only consider mock galaxies of \(\rm scale=0.1kpc\) here. For each galaxy, we also use the bootstrapping method to build 100 galaxies to calculate the corresponding uncertainties, as is shown in Paper I.

{Fig. \ref{fig2} shows the mass distribution information of our mock galaxy samples from TNG50, and the \(\eta-\Sigma\) relation among the mock galaxies. We note that the dark matter here is calculated by \(f_{\rm DM}=1-M_{\rm \star}(<R_{\rm e})/M_{\rm TNG}(<R_{\rm e})\) within three-dimensional \(R_{\rm e}\), to keep consistent with our Paper I. In another aspect, the dark matter fraction varies smoothly along the relation for most galaxies that lower fluctuation {degrees} are associated with higher density and lower dark matter fractions, while higher fluctuation {degrees} are associated with lower density and higher dark matter fractions. The related trends of these parameters are also similar to those of the observational galaxies according to Paper I. However, there exists three dark matter deficient mock galaxies of \(f_{\rm DM}=0\) in the right panel of Fig. \ref{fig2}. Although specifically discussing the dark matter deficient galaxies is beyond the scope of this article, it appears that these galaxies also lie in the \(\eta-\Sigma\) relation at least for our mock sample galaxies from TNG50.}
\section{measurement method}
{It is generally considered that complex dynamical models have higher accuracy in measuring dynamical mass than the application of the virial theorem, but the former are more time consuming. The virial mass estimator derived from the virial theorem is considered to be quick, unbiased and reliable method to derive the dynamical mass of galaxies \citep{2006MNRAS.366.1126C,2022ApJ...936....9V}. Therefore in this paper, we aim to apply the virial theorem to test the influence of \(\eta\) on the measurement of dynamical mass of galaxies. If \(\eta\) does have a certain influence, then we plan to use a complex dynamical model such as Schwarzschild model or JAM to more accurately quantify the degree of influence in our future research.}

The virial dynamical mass estimator of \citet{2006MNRAS.366.1126C} has the form
\begin{equation}\label{func2}
	M_{\rm vir,tot}=\frac{\beta R_{\rm e}\sigma_{\rm e}^2}{G}
\end{equation}
Here, \(\sigma_{\rm e}\) represents the overall velocity dispersion of the line-of-sight direction measured within the circular aperture of \(R_{\rm e}\) radius and \(G\) is the gravitational constant. {TNG galaxies have a parameter named \(R_{\rm half}\), which represents the three-dimensional radius of half-stellar mass of the galaxy. However, compared to three-dimensional radius, two-dimensional half-light radius is more easily obtained from observations. Thus in this paper, we do not use \(R_{\rm half}\), but instead adopting the multi-Gaussian expansion \citep[MGE;][]{1994A&A...285..723E,2002MNRAS.333..400C} to measure two-dimensional \(R_{\rm e}\) of the TNG mock IFU {{stellar}} mass distribution cube, in order to maintain consistency with observations and our Paper I.} The best-fitting scaling factor of \citet{2006MNRAS.366.1126C} is \(\beta=5.0\pm0.1\). In addition, \(\beta\) can also be fitted under the {S\'ersic} model, and they also obtained the fitting equation between {\(\beta\)} and {S\'ersic} index \(n\) as \(\beta(n)=8.87-0.831n+0.0241n^2\). These equations can be used to roughly estimate the dynamical mass, which is efficient and relatively reliable for fast statistics of large samples, and this method is also used in the SAMI DR2 \citep{2018MNRAS.481.2299S}. However, \citet{2006MNRAS.366.1126C} also emphasize that the expression of {\(\beta\)} with respect to {\(n\)} does not improve the estimation of virial mass. Therefore, it seems unnecessary to construct a {S\'ersic} model to measure the {\(n\)} index for each galaxy of the large TNG or even bootstrapping sample. We decided not to introduce the consideration of the {S\'ersic} model. But we cannot simply adopt \(\beta=5.0\), because it is the fitting results of massive galaxies. Moreover, the simulated galaxies of the TNG data may be somehow different from the real galaxies, so we need to fit the TNG mock galaxies of our sample with the equation shape similar to the virial Equation (\ref{func2}).

On the right side of the Equation (\ref{func2}), except for the fitting factor \(\beta\), the remaining items are determined by the morphology (\(R_{\rm e}\)) and kinematics (\(\sigma_{\rm e}\)) of the galaxy. For convenience, we call it \(M_{\rm x}\) here, so we have
\begin{equation}\label{func3}
	M_{\rm x}=\frac{R_{\rm e}\sigma_{\rm e}^2}{G}
\end{equation}
Therefore, what we need is to fit the true mass \(M_{\rm TNG}\) and corresponding \(M_{\rm x}\) inside the mock galaxies. Since {the inner regions of galaxies are generally brighter than the outer regions, the measurement uncertainties of velocity fields in inner regions are relatively small. In addition, the interior of galaxies is generally in a more dynamically stable state,} we also mainly consider the mass fitting within \(R_{\rm e}\) here. And the true mass \(M_{\rm TNG}\) we adopt here is counted by summing the masses of all the gas, stellar, dark matter particles within the three-dimensional \(R_{\rm e}\) to the mock galaxy centres, that is, inside a spherical isosurface of volume \(V=4\pi R_{\rm e}^3/3\). \(\sigma_{\rm e}\) is generally measured from spectrum fitting of \(R_{\rm e}\) aperture spectrum. In the IFS surveys, the \(R_{\rm e}\) aperture spectrum is obtained by co-adding all luminosity-weighted spectrum inside \(R_{\rm e}\). We also used the aperture \(\sigma_{\rm e}\) from SAMI survey to calculate the corresponding virial mass in Paper I. In this work, however, the TNG has no aperture, and the main data of TNG are particles of different masses. Therefore, we use the mass-weighted {stellar} \(V_{\rm rms}\) of each {spaxel of mock IFU data cube} within \(R_{\rm e}\) in every TNG mock galaxy to calculate \(\sigma_{\rm e}\) {of stellar kinematics}, thus \(\sigma_{\rm e}=\sqrt{\langle V_{\rm rms}^2(<R_{\rm e})\rangle}\) \citep{2017ApJ...838...77L}. We use the standard deviation of value between bootstrapping and original mock galaxy to represent the uncertainty of \(\eta\), \(M_{\rm TNG}\), and \(\sigma_{\rm e}\). According to the results of \citet{2013MNRAS.432.1709C} and \citet{2021MNRAS.504.5098D}, {{the uncertainty of \(R_{\rm e}\) obtained by MGE method is around 10 percent for observational data, while according to \citet{2022MNRAS.511.2544D}, the typical uncertainties of semimajor axis \(R_{\rm e,maj}\) is around 7 percent for their simulated galaxies. Therefore, we here set the uncertainty of \(R_{\rm e}\) of our mock galaxies obtained by MGE method to be 10 percent.}} {In addition, because \(\eta\) is the stellar mass-weighted average of \(\eta_{\rm i}\) within \(R_{\rm e}\), thus uncertainties of \(R_{\rm e}\) has little influence on the overall \(\eta\).} The uncertainty of \(M_{\rm x}\) and other values are obtained by error propagation. 

The three-dimensional half-light radius \(r_{\rm 1/2}\) and the corresponding \(M_{\rm 1/2}\) (mass within the sphere of radius \(r_{\rm 1/2}\) enclosing half of the total galaxy light) {are also usually involved in galaxy dynamical models \citep{2010MNRAS.406.1220W,2013MNRAS.432.1709C}}. It should be noted {again} that \(R_{\rm e}\) is {generally} the two-dimensional half-light radius after projection. Thus the \(M_{\rm TNG}\) here is not equal to \(M_{\rm 1/2}\). However, on the one hand, from the perspective of observation, \(R_{\rm e}\) is a relatively more observational quantity, which is easy to obtain from observational images. On the other hand, the main purpose of our article is to examine whether the dynamical mass fitting deviation varies with \(\eta\) in the inner part of galaxies. Therefore, there is no essential difference when we use \(M_{\rm TNG}\) and \(R_{\rm e}\) instead of \(M_{\rm 1/2}\) and \(r_{\rm 1/2}\).

Due to the relatively large range of masses of galaxies, it is better to do a linear fitting in logarithmic scale. General linear fitting requires two factors, that is, fitting \(y=a+bx\). On the other hand, we also want to obtain the fitting equation of a single factor similar to Equation (\ref{func2}) for comparison with other research works. So we use both two approaches here, we call them \(M_{\rm vir1}\) for linear fitting {with fixed slope} and \(M_{\rm vir2}\) for {general linear fitting} respectively. That is, fitting \(\log M_{\rm TNG}\) with the equation \(\log M_{\rm vir1}=\log \beta_{\rm TNG,re}+\log M_{\rm x}\), and \(\log M_{\rm vir2}=a+b\log M_{\rm x}\). The {general linear fitting} is not purely in the mathematical sense, factor \(b\) can describe the deviation of real galaxies relative to the assumption of virial theorem. It needs to be emphasized again that the virial mass \(M_{\rm vir}\) here refers to the estimated dynamical mass within the three-dimensional \(R_{\rm e}\) range to the galaxy centre, so it has a different meaning from the virial total mass \(M_{\rm vir,tot}\) in Equation (\ref{func2}).

{{We adopt the free implementation {\tiny LTS\_LINEFIT} of}  \citet{2013MNRAS.432.1709C} {for both linear fits.}} {\tiny LTS\_LINEFIT} adopt the FAST\_LTS method developed by \citet{ROUSSEEUW2006} to apply least trimmed squares (LTS) regression. The {\tiny LTS\_LINEFIT} algorithm can iteratively remove outliers and is a very reliable and robust linear fitting method. It can also be used to obtain the systematic difference of different measurement methods \citep{2021MNRAS.504.5098D}. {{In this work, we adopt the \(2.6\sigma\) clipping, which means that data points deviate more than \(2.6\sigma\) from the best fit are considered to be outliers and are excluded during linear fitting, while 99 percent of the data points would be included within \(2.6\sigma\) region for a Gaussian distribution.}}
\section{results}
\begin{figure}
	\begin{center}
		\includegraphics[width=0.45\textwidth,angle=0]{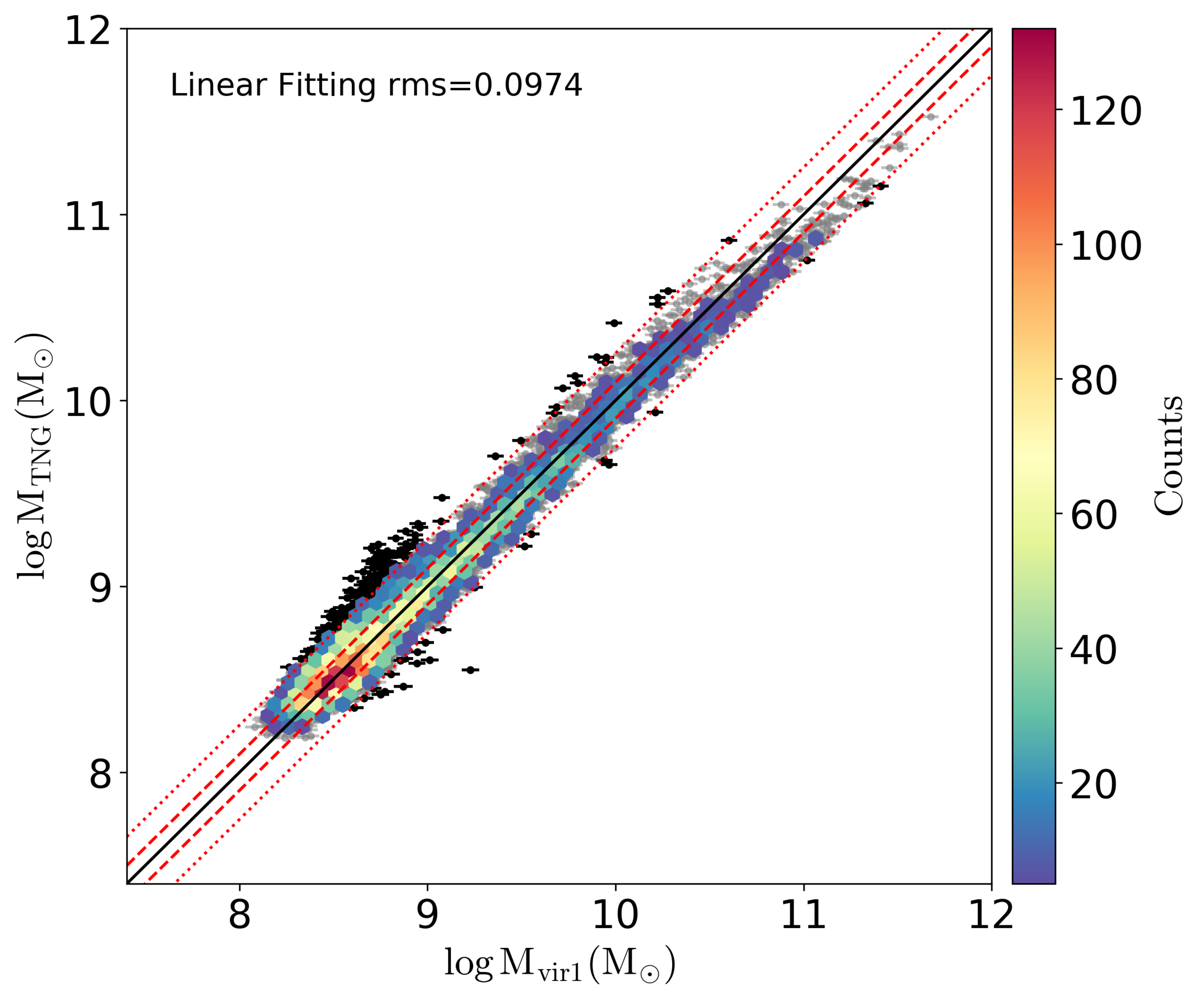}
	\end{center}
	\caption{The linear fitting {with fixed slope} of \(\log M_{\rm vir1}\). The solid line represents one-to-one relation, while the two dashed lines indicate \(1\sigma\) (containing 68 percent values under the Gaussian distribution assumption) region and two dotted lines indicate \(2.6\sigma\) (99 percent) region. Here \(\sigma\) means linear fitting rms of fitting deviation \(\log M_{\rm vir1}-\log M_{\rm TNG}\) and \(\sigma_{\rm vir1}=0.0974\). {{The black data points being outside of \(2.6\sigma\) region were considered outliers and were removed during the fitting process, while the grey points show the mock galaxies involved in the linear fit.}} {Due to the overlap of sample points, the colourbar displays the number counts of galaxies within the corresponding hexagonal bin.}
	}\label{fig3}
\end{figure}

\begin{figure}
	\begin{center}
		\includegraphics[width=0.45\textwidth,angle=0]{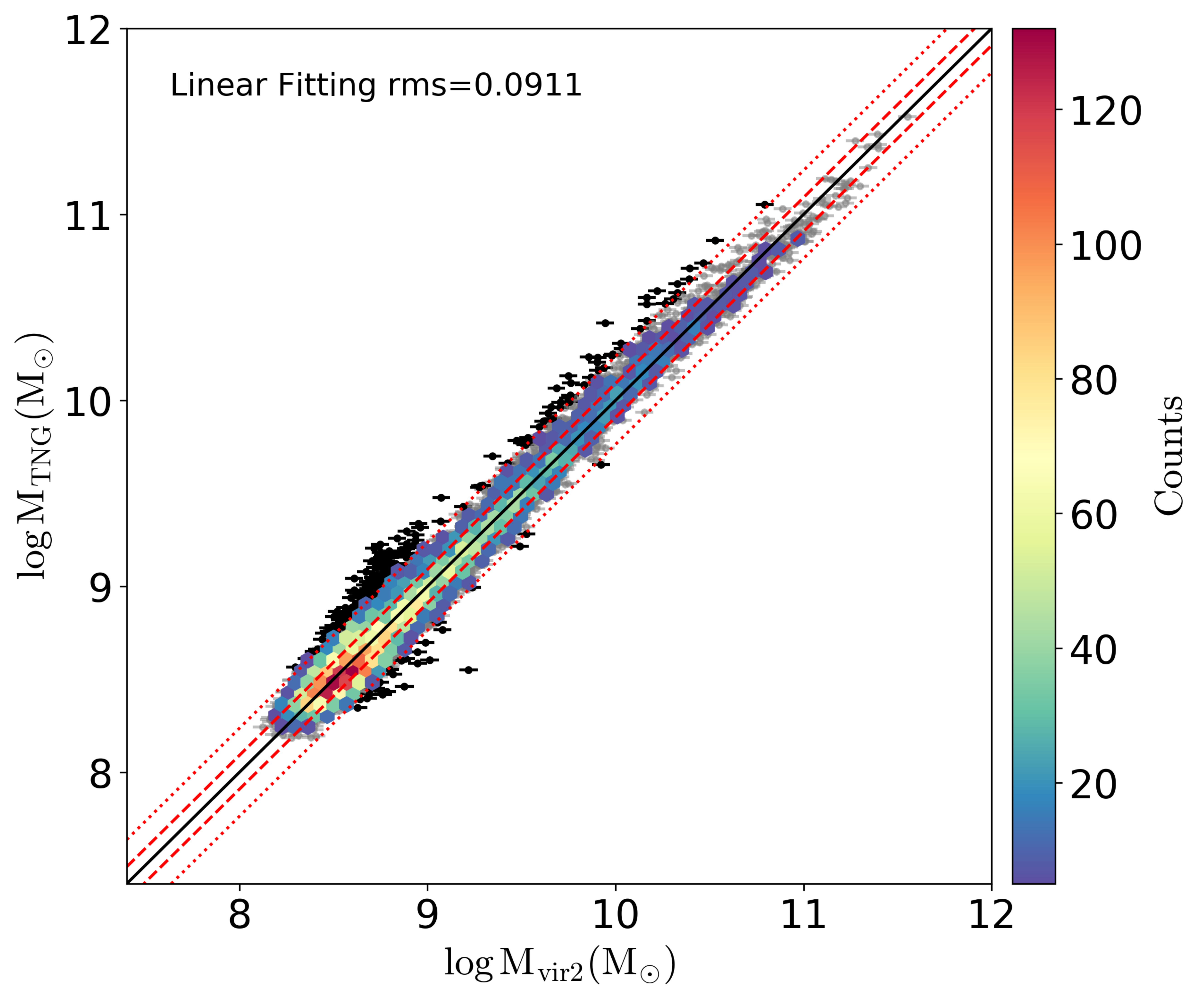}
	\end{center}
	\caption{The {general linear fitting} of \(\log M_{\rm vir2}\). The meaning of the lines and hexagonal bins is the same as in Fig. \ref{fig3}, and the linear fitting rms is \(\sigma_{\rm vir2}=0.0911\). 
	}\label{fig4}
\end{figure}

Our linear fitting results for linear fitting {with fixed slope} and {general linear fitting} are as follows, respectively,
\begin{equation}\label{func4}
\log M_{\rm vir1}=(0.5954\pm0.0012)+\log M_{\rm x}
\end{equation}
\begin{equation}\label{func5}
\log M_{\rm vir2}=(0.9818\pm0.0155)+(0.9540\pm0.0018)\log M_{\rm x}
\end{equation}
The fitting lines are also shown in Fig. \ref{fig3} and Fig. \ref{fig4}. {{The hexagonal bins are used to show the overlapped sample points.}} The linear fitting rms of fitting deviation \(\log M_{\rm vir}-\log M_{\rm TNG}\) are \(\sigma_{\rm vir1}=0.0974\) and \(\sigma_{\rm vir2}=0.0911\). This is also reasonable {{that}} {general linear fitting} should theoretically be better than linear fitting {with fixed slope}. {We note that the uncertainties of linear fitting parameters being very small should come from the large sample size. In addition, the linear fitting rms are small for both fitting lines}. It can be said that, at least in TNG simulations, virial dynamical mass estimator is still very reliable and robust {{within about 0.1 dex for most mock galaxies over a wide mass range}}.

\begin{table}
	\large
	\caption{{{Local mass estimation bias of each even bin of \(\log\eta\).}} \label{table1}} 
	\begin{center}
		\begin{tabular}{|c|c|c|}
			\hline
			Middle \(\log\eta\) of the bin & \(\langle\Delta\log M_{\rm vir1}\rangle_{\rm bin}\) &  \(\langle\Delta\log M_{\rm vir2}\rangle_{\rm bin}\) \\
			\hline
            -1.3923 & 0.0254 & -0.0659 \\
            -1.2883 & 0.0948 & 0.0053 \\
            -1.1844 & 0.1020 & 0.0302 \\
            -1.0805 & 0.1067 & 0.0344 \\
            -0.9765 & 0.0700 & 0.0084 \\
            -0.8726 & 0.0498 & -0.0041 \\
            -0.7686 & 0.0119 & -0.0315 \\
            -0.6647 & -0.0125 & -0.0395 \\
            -0.5608 & -0.0143 & -0.0318 \\
            -0.4568 & -0.0368 & -0.0452 \\
            -0.3529 & -0.0204 & -0.0237 \\
            -0.2489 & -0.0180 & -0.0123 \\
            -0.1450 & -0.0072 & 0.0082 \\
            -0.0411 & 0.0328 & 0.0530 \\
            0.0629 & 0.0817 & 0.1006 \\
			\hline
		\end{tabular}
	\end{center}
\end{table}

\begin{figure}
	\begin{center}
		\includegraphics[width=0.45\textwidth,angle=0]{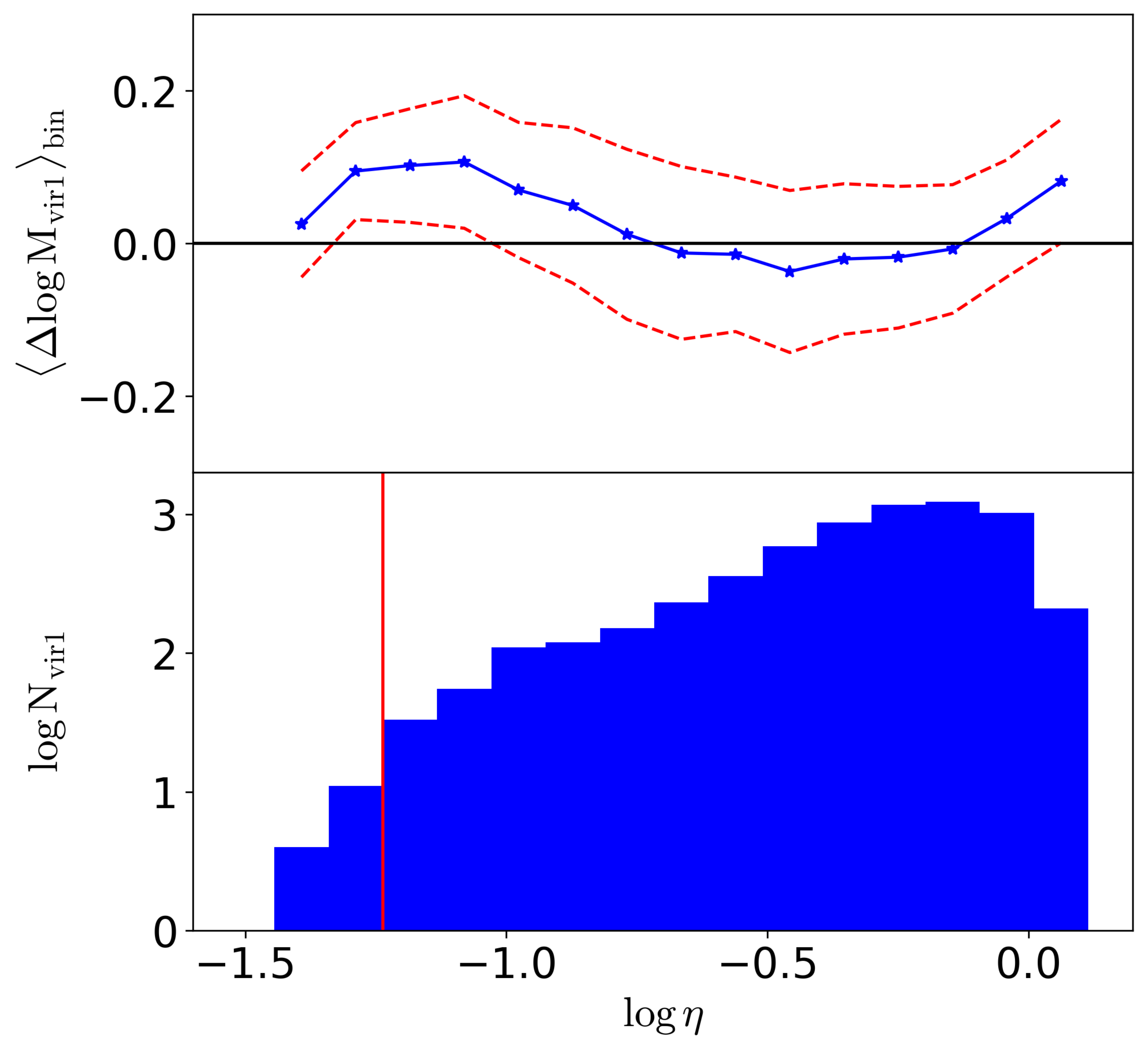}
	\end{center}
	\caption{{{Local bias for the overall virial estimator of \(\log M_{\rm vir1}\). Upper panel: The distribution of local mass estimation bias \(\langle\Delta\log M_{\rm vir1}\rangle_{\rm bin}\) in each even bin of \(\log\eta\). The dashed lines show the standard deviation of mass estimation deviation relative to the local bias of each bin. Lower panel: Distribution histogram of the number of mock galaxies \(\log N_{\rm vir1}\) in each bin. The sample numbers in the two bins on the left side of the vertical straight line are too small, so they are not included in the subsequent random sample selection and calculations.}}
	}\label{fig5}
\end{figure}

\begin{figure}
	\begin{center}
		\includegraphics[width=0.45\textwidth,angle=0]{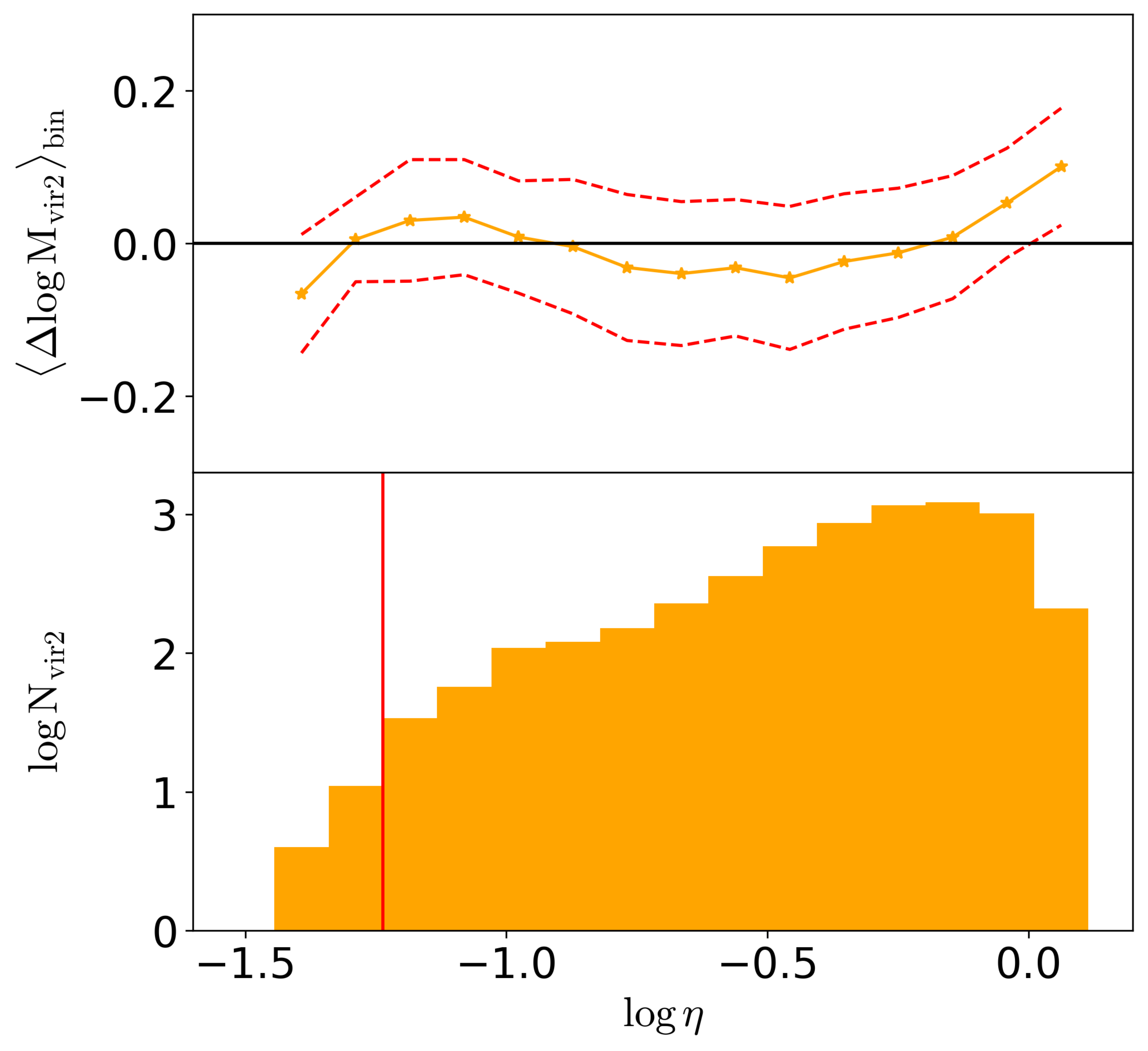}
	\end{center}
	\caption{{{Local bias for the overall virial estimator of \(\log M_{\rm vir2}\). The meaning of the lines in the two panels are the same as in Fig. \ref{fig5}. The numbers in each bin are similar to those in Fig. \ref{fig5}.
		}}
	}\label{fig6}
\end{figure}

{{Although the average mass estimation deviation of all mock galaxies, that is, the total bias is nearly zero, the sample contains different types of galaxies, such as ETGs or spiral galaxies that have essentially different mass distributions, so there may be some bias if we apply the virial estimators only to galaxies of similar type. Since we mainly focus on the kinematical asymmetry of galaxies in this paper, here we use \(\eta\) to distinguish the types of galaxies and then try to study the local bias of different galaxy types. The relative uncertainty of \(\eta\) is approximately 25\% by bootstrapping, thus we divide the sample mock galaxies to 15 bins evenly according to the \(\log\eta\) range, and the \(\log\eta\) width of each bin is approximately 0.1 dex. Within each bin, the samples can be considered to have the same degree of kinematical fluctuations. The local average mass estimation biases \(\langle\Delta\log M_{\rm vir}\rangle_{\rm bin}\) of each bin for the two linear fits are shown in Table \ref{table1} and are displayed in the upper panels of Fig. \ref{fig5} and Fig. \ref{fig6}, while \(\Delta\log M_{\rm vir}=\log M_{\rm vir}-\log M_{\rm TNG}\) represents the mass deviation. The lower panels of Fig. \ref{fig5} and Fig. \ref{fig6} show the mock galaxy numbers involved in the two linear fits in each bin. The leftmost two bins have too few data points, which may significantly increase the uncertainty of the bin bias, so we mark them with red lines and remove them in the subsequent calculations.}}

{{As shown in left panel of fig. 2 of Paper I, \(\eta\) is negatively correlated with the galaxy stellar mass, and according to the tight \(\eta-\Sigma\) relation, \(\eta\) is log-linearly negatively correlated with the galaxy stellar surface density. Therefore, distinguishing galaxy types according to \(\log\eta\) is also equivalent to distinguishing galaxy types according to stellar mass (non-linearly) or surface density (log-linearly). Mock galaxies of the bins for the minimum \(\log\eta\) should be dominated by massive galaxies, which are usually ETGs, and the local positive bias indicates that their masses are overestimated by the overall virial estimator; similarly, galaxies of the bin of \(\log\eta \sim -0.5\) approximately correspond to intermediate-mass galaxies, generally dominated by spiral galaxies, and the negative bias indicates the underestimated dynamical masses. Fig.1 of}} \citet{2022ApJ...936....9V} {{also shows that applying the virial mass estimator obtained by ETGs to spiral galaxies will underestimate their dynamical mass, which is consistent with our results here. The maximum bias deviation of the two bins in Table \ref{table1} is about 0.16 dex; we think this may indicate the systematic difference in the virial mass estimation of ETGs and spiral galaxies. Overall, these local biases can help the virial mass estimators obtained from observations of massive galaxies, such as those of}} \citet{2006MNRAS.366.1126C}{{, to be applied to intermediate-mass or low-mass galaxies.}}

We can derive \(M_{\rm vir1}=(3.9389\pm0.0113)\,M_{\rm x}\) from Equation (\ref{func4}), so the scaling factor from \(M_{\rm x}\) to \(M_{\rm vir1}\) is \(\beta_{\rm TNG,re}=3.9389\pm0.0113\). However, since \(M_{\rm vir1}\) is within the range of three-dimensional \(R_{\rm e}\), and generally there is \(r_{\rm 1/2}\approx 1.33R_{\rm e}\) \citep{2013MNRAS.432.1709C}{{, in order to}} roughly compare the scaling factor of \(\beta_{\rm TNG,re}\) with those from other research works, we simply assume that the dynamical mass space density in the galaxy inner region is near constant, and then \(M_{\rm vir1,1/2}\approx 1.33^3M_{\rm vir1}\approx9.27\,M_{\rm x}\). On the other hand, we can obtain \(M_{\rm vir,1/2}=2.5\,M_{\rm x}\) from Equation (\ref{func2}) for \citet{2006MNRAS.366.1126C}, and in \citet{2010MNRAS.406.1220W} they obtained \(M_{\rm 1/2}\simeq4.0M_{\rm x}\). It can be seen that \(\beta_{\rm TNG}\) is quite large compared with the observational results, roughly twice as much as in other research works. {One possible reason is that here we are using both star-forming and non-star-forming galaxies, but \citet{2006MNRAS.366.1126C} and \citet{2010MNRAS.406.1220W} mainly focus on ETGs, which are to first order non-star-forming. {{Table \ref{table1} can also prove that the scaling factor corresponding to ETGs is relatively small}}. Another possible reason is that it may} due to the much greater proportion of dark matter inside \(R_{\rm e}\) of TNG galaxies. \citet{2018MNRAS.481.1950L} measured the dark matter fractions within TNG galaxies. In their fig. 12, the dark matter fractions of TNG galaxy within \(R_{\rm e}\) are quite larger than the observational results (more than twice the measurement result of \citealt{2013MNRAS.432.1709C}), but when the measurement radius is enlarged, from the distance of \(5R_{\rm e}\) or even farther, the difference of dark matter fraction between TNG galaxies and actual galaxies is not so large as described in \citet{2018MNRAS.481.1950L}. This shows that at least within \(R_{\rm e}\), the mass distribution of TNG galaxies is somewhat different from that of actual galaxies. The difference in the scale factors only shows their different mass distribution within \(R_{\rm e}\), and does not mean that the fitting result of the virial mass estimator is not good. In addition, in Paper I, we have showed that TNG galaxies can still perfectly display the relation of \(\eta\) with stellar surface mass density very similar to that from observation, so we think that the difference in scaling factor \(\beta\) will not cause too much problem.

\begin{table}
	\large
	\caption{Spearman's rank correlation coefficients between \(\eta\) and mass deviations {{relative to local bias.}} \label{table2}} 
	\begin{center}
		\begin{tabular}{|c|c|c|}
			\hline
			\(\log\eta\) versus intrinsic mass deviations & Spearman's \(\rho\) & \(p\)-value  \\
			\hline
			\(\left|\Delta\log M_{\rm vir1}-\langle\Delta\log M_{\rm vir1}\rangle_{\rm bin}\right|\) & \(-0.0127\) & \(0.3074\)  \\
			
			\(\left|\Delta\log M_{\rm vir2}-\langle\Delta\log M_{\rm vir2}\rangle_{\rm bin}\right|\) & \(0.0064\) & \(0.6073\)   \\
			
			\(\Delta\log M_{\rm vir1}-\langle\Delta\log M_{\rm vir1}\rangle_{\rm bin}\) & \(0.0011\) & \(0.9304\) \\
			
			\(\Delta\log M_{\rm vir2}-\langle\Delta\log M_{\rm vir2}\rangle_{\rm bin}\) & \(0.0057\) & \(0.6449\)  \\
			\hline
		\end{tabular}
	\end{center}
\end{table}

\begin{figure}
	\begin{center}
		\includegraphics[width=0.45\textwidth,angle=0]{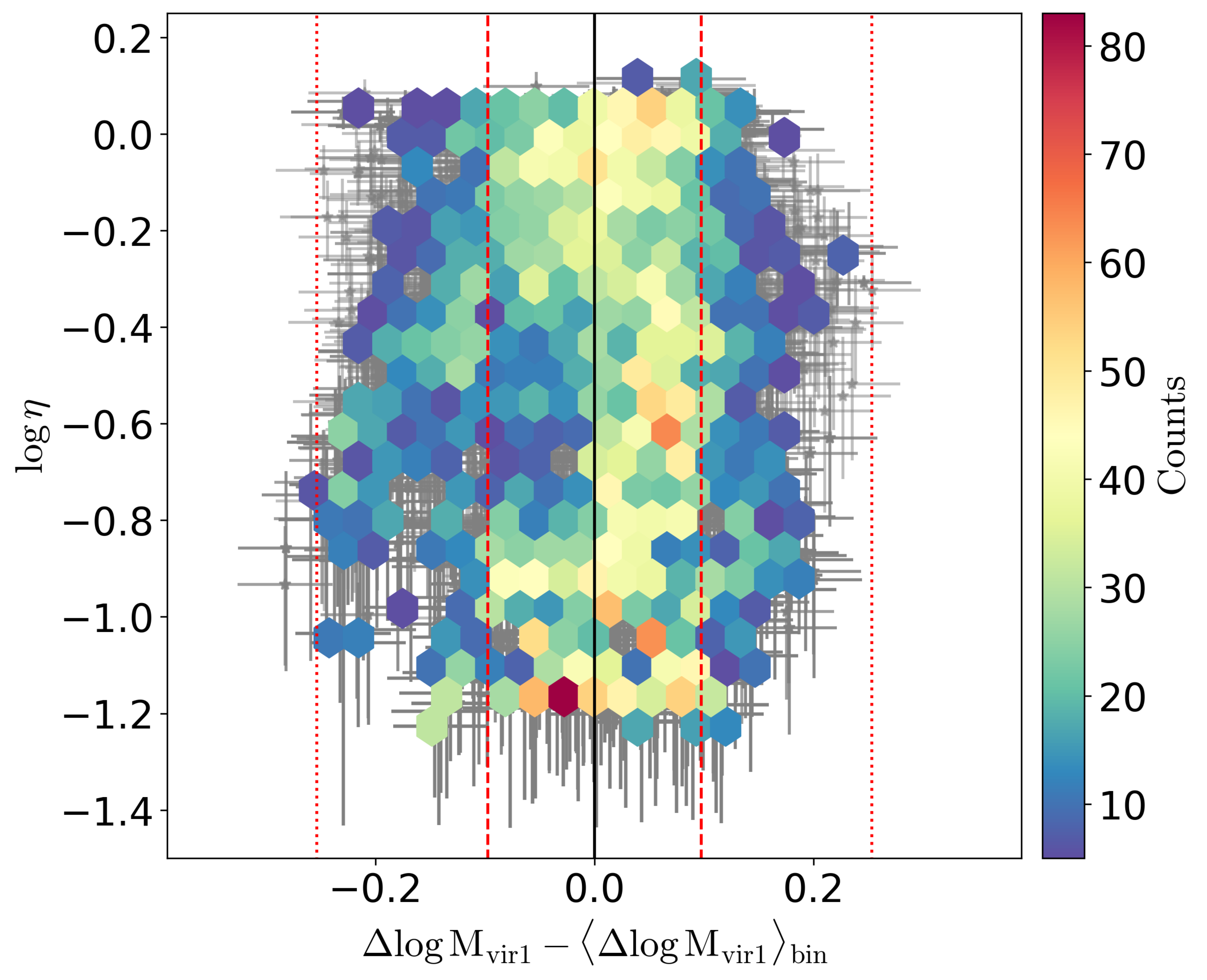}
	\end{center}
	\caption{{{The distribution of linear fitting deviation relative to the local bias \(\Delta\log M_{\rm vir1}-\langle\Delta\log M_{\rm vir1}\rangle_{\rm bin}\) with the asymmetry parameter \(\eta\) for the randomly selected sample. Vertical lines show the same levels of \(\sigma_{\rm vir1}\) width as in Fig. \ref{fig3} and hexagonal bins also show the number of overlapped random sample points. This figure shows the uncorrelation of \(\eta\) and the intrinsic mass estimation deviation if the biases can be eliminated properly}}.
	}\label{fig7}
\end{figure}

\begin{figure}
	\begin{center}
		\includegraphics[width=0.45\textwidth,angle=0]{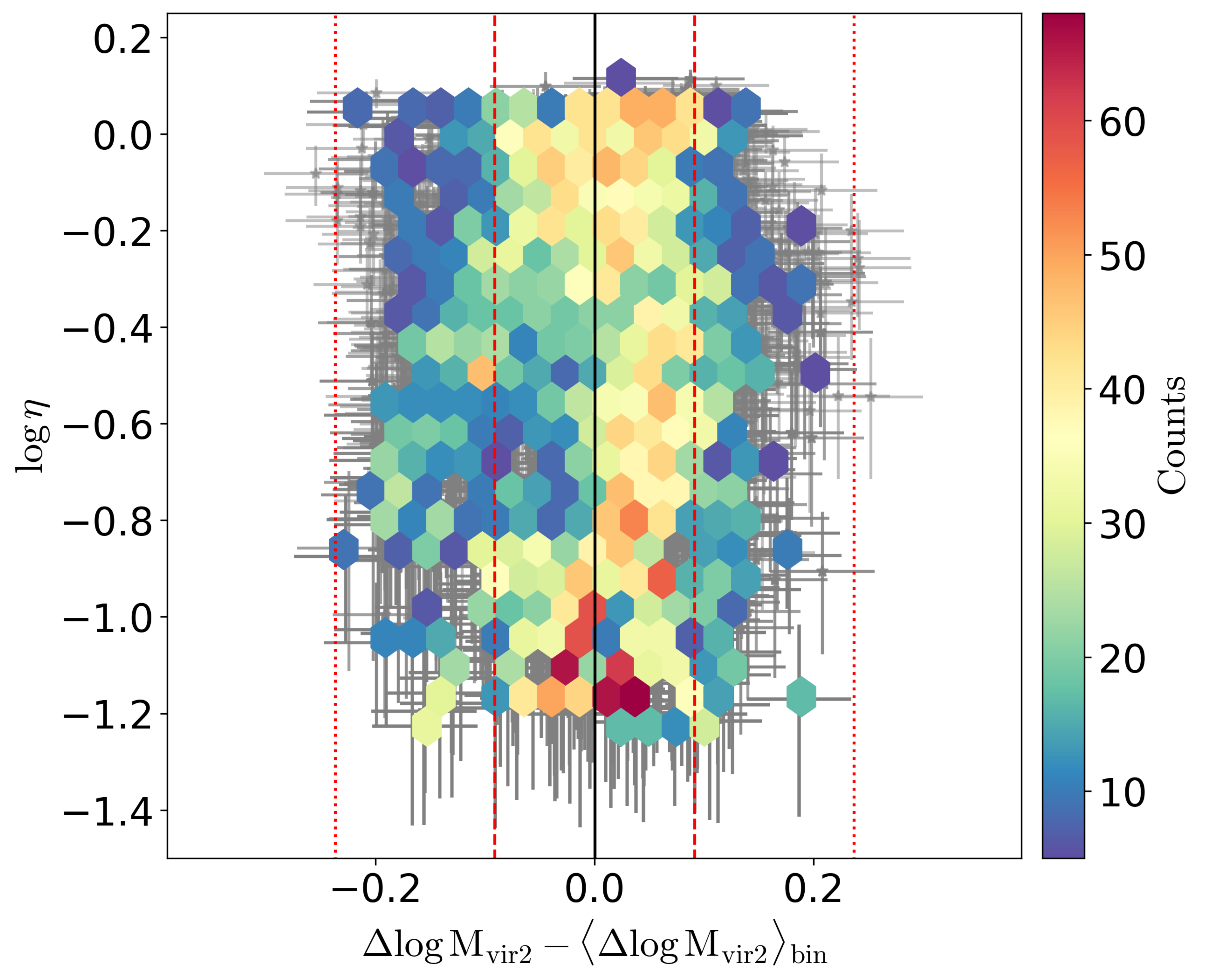}
	\end{center}
	\caption{{{The distribution of linear fitting deviation relative to the local bias \(\Delta\log M_{\rm vir2}-\langle\Delta\log M_{\rm vir2}\rangle_{\rm bin}\) with the asymmetry parameter \(\eta\) for the randomly selected sample. Vertical lines show the same levels of \(\sigma_{\rm vir2}\) width as in Fig. \ref{fig4} and hexagonal bins also show the number of overlapped random sample points. \(\eta\) is also uncorrelated with the intrinsic mass estimation deviation for \(\log M_{\rm vir2}\)}}.
	}\label{fig8}
\end{figure}

{{It should be noted that the local bias is not caused by kinematical fluctuations of the galaxy. This is because if we only use the corresponding data points in each bin to fit the virial mass estimators, it can be expected that the local bias will be eliminated. Therefore, we need to study whether kinematical fluctuations significantly change the intrinsic mass estimation deviations, that is, compare the relationship between \(\log\eta\) and mass estimation deviations relative to the local bias \(\Delta\log M_{\rm vir}-\langle\Delta\log M_{\rm vir}\rangle_{\rm bin}\). Since the number of galaxies in each bin is different, we repeatedly randomly selected 500 mock galaxies from the 13 groups of bins on the right side of the red line in Fig. \ref{fig5} and Fig. \ref{fig6} to construct the new random samples for investigation.}} The results are shown in Fig. \ref{fig7} and Fig. \ref{fig8}, respectively. {The hexagonal bins are used to show the overlapped sample points.} The {{average}} deviation width on the \(x\)-axis for the similar \(\eta\) shows the {{intrinsic virial dynamical mass estimation deviations}}. Therefore, if the fluctuation degree really affects the accuracy of dynamical mass measurement, then we will obtain an image similar to an inverted triangle or an inverted cone in Fig. \ref{fig7} and Fig. \ref{fig8}. However, as shown in these figures, the deviation width of the \(x\)-axis does not change significantly with \(\eta\). {{Moreover, we use the Spearman's rank correlation coefficient to quantify the correlations. The results are shown in Table \ref{table2}. The Spearman's \(\rho\) values of the two \(\Delta\log M_{\rm vir}-\langle\Delta\log M_{\rm vir}\rangle_{\rm bin}\) show that the local biases are eliminated properly, while Spearman's \(\rho\) values of the two \(\left|\Delta\log M_{\rm vir}-\langle\Delta\log M_{\rm vir}\rangle_{\rm bin}\right|\) could show that there is no correlation between \(\eta\) and intrinsic mass deviations. In other words, regardless of the kinematical fluctuation degree of galaxies, the intrinsic mass deviations do not change significantly when applying the virial mass estimations.}} Again, Paper I shows that there is high degree of kinematical small-scale fluctuations in the stellar motion of low-mass galaxies of TNG50, which is consistent with the SAMI survey observations. So the results here can also {indicate} that the existence of small-scale fluctuations does not affect or bend the virial theorem of low-mass galaxies, nor does it affect the measurement results of the dynamical masses by the virial dynamical mass {estimator}.

\section{summary}
The widely used dynamical models all assume that the {stellar kinematics} of the galaxy is two-dimensional centrosymmetric after projection; however, the existence of kinematical small-scale fluctuations {in the stellar velocity fields} will cause the actual stellar motion to be unsmooth and deviate from symmetric pattern in galaxies. So we want to test whether the increasing small-scale fluctuation degree will increase the deviation degree of the general dynamical mass estimation. We use the asymmetry parameter \(\eta\) to quantify the kinematical small-scale fluctuation degree, and \(\eta\) is derived from applying the \(180^{\circ}\) rotation self-subtraction method to the second moment \(V_{\rm rms}\) velocity fields. We used the TNG50 numerical simulation data to construct mock galaxies, and used the virial mass estimator to fit the true total mass inside the sphere of radius \(R_{\rm e}\), which is the two-dimensional circularized half-stellar mass radius of mock galaxies. The difference between the fitting factor and those from the observational results shows that the mass distribution of TNG galaxies is different from the actual galaxies inside \(R_{\rm e}\), but the fitting results with the virial mass estimator have small scatter and uncertainties. Our results show that the form of the virial dynamical mass estimator is overall relatively reliable, and for the mock galaxies in this work, the dynamical masses of \(\log M_{\rm vir}\) calculated with the symmetric assumption are still accurate within around 0.1 dex. We divided these mock galaxies into 15 bins evenly according to the \(\log\eta\) range to study the local bias of virial mass estimation. Our results show that the maximum bias difference between bins is about 0.16 dex, which we believe may come from the systematic difference between ETGs and spiral galaxies. The local biases we obtained can also help the virial mass estimators obtained from observations of massive galaxies to be applied to galaxies of other masses. In addition, the intrinsic mass deviation estimated from virial estimator has no relation with the asymmetry parameter \(\eta\) of galaxies if the bias is eliminated properly, indicating that the kinematical small-scale fluctuations of galaxies do not affect the estimation of the overall dynamical mass of galaxies, even for low-mass galaxies. This also indicates that the kinematical small-scale fluctuations seems to have less effect on the overall galaxy dynamics.

\section*{Acknowledgements}
We are grateful to the anonymous referee for the valuable comments that improved the quality of this paper. This study was supported by the National Natural Science Foundation of China under grant nos. 11988101, and 11890694, and the National Key R\&D Program of China (nos. 2019YFA0405500).
\section*{Data Availability}
The TNG50 data of IllustrisTNG project used for this study are publicly available at https://www.tng-project.org.
\bibliographystyle{mnras}
\bibliography{kineflucdynmass}

\begin{thebibliography}{}
\makeatletter
\relax
\def\mn@urlcharsother{\let\do\@makeother \do\$\do\&\do\#\do\^\do\_\do\%\do\~}
\def\mn@doi{\begingroup\mn@urlcharsother \@ifnextchar [ {\mn@doi@}
  {\mn@doi@[]}}
\def\mn@doi@[#1]#2{\def\@tempa{#1}\ifx\@tempa\@empty \href
  {http://dx.doi.org/#2} {doi:#2}\else \href {http://dx.doi.org/#2} {#1}\fi
  \endgroup}
\def\mn@eprint#1#2{\mn@eprint@#1:#2::\@nil}
\def\mn@eprint@arXiv#1{\href {http://arxiv.org/abs/#1} {{\tt arXiv:#1}}}
\def\mn@eprint@dblp#1{\href {http://dblp.uni-trier.de/rec/bibtex/#1.xml}
  {dblp:#1}}
\def\mn@eprint@#1:#2:#3:#4\@nil{\def\@tempa {#1}\def\@tempb {#2}\def\@tempc
  {#3}\ifx \@tempc \@empty \let \@tempc \@tempb \let \@tempb \@tempa \fi \ifx
  \@tempb \@empty \def\@tempb {arXiv}\fi \@ifundefined
  {mn@eprint@\@tempb}{\@tempb:\@tempc}{\expandafter \expandafter \csname
  mn@eprint@\@tempb\endcsname \expandafter{\@tempc}}}

\bibitem[\protect\citeauthoryear{{Bell} \& {de Jong}}{{Bell} \& {de
  Jong}}{2001}]{2001ApJ...550..212B}
{Bell} E.~F.,  {de Jong} R.~S.,  2001, \mn@doi [\apj] {10.1086/319728}, \href
  {https://ui.adsabs.harvard.edu/abs/2001ApJ...550..212B} {550, 212}

\bibitem[\protect\citeauthoryear{{Bell}, {McIntosh}, {Katz}  \&
  {Weinberg}}{{Bell} et~al.}{2003}]{2003ApJS..149..289B}
{Bell} E.~F.,  {McIntosh} D.~H.,  {Katz} N.,   {Weinberg} M.~D.,  2003, \mn@doi
  [\apjs] {10.1086/378847}, \href
  {https://ui.adsabs.harvard.edu/abs/2003ApJS..149..289B} {149, 289}

\bibitem[\protect\citeauthoryear{{Binney} \& {Tremaine}}{{Binney} \&
  {Tremaine}}{2008}]{2008gady.book.....B}
{Binney} J.,  {Tremaine} S.,  2008, {Galactic Dynamics: Second Edition}

\bibitem[\protect\citeauthoryear{{Bloom} et~al.,}{{Bloom}
  et~al.}{2017}]{2017MNRAS.465..123B}
{Bloom} J.~V.,  et~al., 2017, \mn@doi [\mnras] {10.1093/mnras/stw2605}, \href
  {https://ui.adsabs.harvard.edu/abs/2017MNRAS.465..123B} {465, 123}

\bibitem[\protect\citeauthoryear{{Bundy} et~al.,}{{Bundy}
  et~al.}{2015}]{2015ApJ...798....7B}
{Bundy} K.,  et~al., 2015, \mn@doi [\apj] {10.1088/0004-637X/798/1/7}, \href
  {https://ui.adsabs.harvard.edu/abs/2015ApJ...798....7B} {798, 7}

\bibitem[\protect\citeauthoryear{{Cappellari}}{{Cappellari}}{2002}]{2002MNRAS.333..400C}
{Cappellari} M.,  2002, \mn@doi [\mnras] {10.1046/j.1365-8711.2002.05412.x},
  \href {https://ui.adsabs.harvard.edu/abs/2002MNRAS.333..400C} {333, 400}

\bibitem[\protect\citeauthoryear{{Cappellari}}{{Cappellari}}{2008}]{2008MNRAS.390...71C}
{Cappellari} M.,  2008, \mn@doi [\mnras] {10.1111/j.1365-2966.2008.13754.x},
  \href {https://ui.adsabs.harvard.edu/abs/2008MNRAS.390...71C} {390, 71}

\bibitem[\protect\citeauthoryear{{Cappellari}}{{Cappellari}}{2016}]{2016ARA&A..54..597C}
{Cappellari} M.,  2016, \mn@doi [\araa] {10.1146/annurev-astro-082214-122432},
  \href {https://ui.adsabs.harvard.edu/abs/2016ARA&A..54..597C} {54, 597}

\bibitem[\protect\citeauthoryear{{Cappellari}}{{Cappellari}}{2020}]{2020MNRAS.494.4819C}
{Cappellari} M.,  2020, \mn@doi [\mnras] {10.1093/mnras/staa959}, \href
  {https://ui.adsabs.harvard.edu/abs/2020MNRAS.494.4819C} {494, 4819}

\bibitem[\protect\citeauthoryear{{Cappellari} et~al.,}{{Cappellari}
  et~al.}{2006}]{2006MNRAS.366.1126C}
{Cappellari} M.,  et~al., 2006, \mn@doi [\mnras]
  {10.1111/j.1365-2966.2005.09981.x}, \href
  {https://ui.adsabs.harvard.edu/abs/2006MNRAS.366.1126C} {366, 1126}

\bibitem[\protect\citeauthoryear{{Cappellari} et~al.,}{{Cappellari}
  et~al.}{2011}]{2011MNRAS.413..813C}
{Cappellari} M.,  et~al., 2011, \mn@doi [\mnras]
  {10.1111/j.1365-2966.2010.18174.x}, \href
  {https://ui.adsabs.harvard.edu/abs/2011MNRAS.413..813C} {413, 813}

\bibitem[\protect\citeauthoryear{{Cappellari} et~al.,}{{Cappellari}
  et~al.}{2013}]{2013MNRAS.432.1709C}
{Cappellari} M.,  et~al., 2013, \mn@doi [\mnras] {10.1093/mnras/stt562}, \href
  {https://ui.adsabs.harvard.edu/abs/2013MNRAS.432.1709C} {432, 1709}

\bibitem[\protect\citeauthoryear{{Conselice}}{{Conselice}}{2003}]{2003ApJS..147....1C}
{Conselice} C.~J.,  2003, \mn@doi [\apjs] {10.1086/375001}, \href
  {https://ui.adsabs.harvard.edu/abs/2003ApJS..147....1C} {147, 1}

\bibitem[\protect\citeauthoryear{{Conselice}, {Bershady}  \&
  {Jangren}}{{Conselice} et~al.}{2000}]{2000ApJ...529..886C}
{Conselice} C.~J.,  {Bershady} M.~A.,   {Jangren} A.,  2000, \mn@doi [\apj]
  {10.1086/308300}, \href
  {https://ui.adsabs.harvard.edu/abs/2000ApJ...529..886C} {529, 886}

\bibitem[\protect\citeauthoryear{{Croom} et~al.,}{{Croom}
  et~al.}{2021}]{2021MNRAS.505..991C}
{Croom} S.~M.,  et~al., 2021, \mn@doi [\mnras] {10.1093/mnras/stab229}, \href
  {https://ui.adsabs.harvard.edu/abs/2021MNRAS.505..991C} {505, 991}

\bibitem[\protect\citeauthoryear{{D'Eugenio} et~al.,}{{D'Eugenio}
  et~al.}{2021}]{2021MNRAS.504.5098D}
{D'Eugenio} F.,  et~al., 2021, \mn@doi [\mnras] {10.1093/mnras/stab1146}, \href
  {https://ui.adsabs.harvard.edu/abs/2021MNRAS.504.5098D} {504, 5098}

\bibitem[\protect\citeauthoryear{{Emsellem}, {Monnet}  \& {Bacon}}{{Emsellem}
  et~al.}{1994}]{1994A&A...285..723E}
{Emsellem} E.,  {Monnet} G.,   {Bacon} R.,  1994, \aap, \href
  {https://ui.adsabs.harvard.edu/abs/1994A&A...285..723E} {285, 723}

\bibitem[\protect\citeauthoryear{{Graham} et~al.,}{{Graham}
  et~al.}{2018}]{2018MNRAS.477.4711G}
{Graham} M.~T.,  et~al., 2018, \mn@doi [\mnras] {10.1093/mnras/sty504}, \href
  {https://ui.adsabs.harvard.edu/abs/2018MNRAS.477.4711G} {477, 4711}

\bibitem[\protect\citeauthoryear{{Harborne}, {van de Sande}, {Cortese},
  {Power}, {Robotham}, {Lagos}  \& {Croom}}{{Harborne}
  et~al.}{2020}]{2020MNRAS.497.2018H}
{Harborne} K.~E.,  {van de Sande} J.,  {Cortese} L.,  {Power} C.,  {Robotham}
  A.~S.~G.,  {Lagos} C.~D.~P.,   {Croom} S.,  2020, \mn@doi [\mnras]
  {10.1093/mnras/staa1847}, \href
  {https://ui.adsabs.harvard.edu/abs/2020MNRAS.497.2018H} {497, 2018}

\bibitem[\protect\citeauthoryear{{Kim}, {Ho}, {Peng}, {Barth}  \& {Im}}{{Kim}
  et~al.}{2017}]{2017ApJS..232...21K}
{Kim} M.,  {Ho} L.~C.,  {Peng} C.~Y.,  {Barth} A.~J.,   {Im} M.,  2017, \mn@doi
  [\apjs] {10.3847/1538-4365/aa8a75}, \href
  {https://ui.adsabs.harvard.edu/abs/2017ApJS..232...21K} {232, 21}

\bibitem[\protect\citeauthoryear{{Krajnovi{\'c}}, {Cappellari}, {de Zeeuw}  \&
  {Copin}}{{Krajnovi{\'c}} et~al.}{2006}]{2006MNRAS.366..787K}
{Krajnovi{\'c}} D.,  {Cappellari} M.,  {de Zeeuw} P.~T.,   {Copin} Y.,  2006,
  \mn@doi [\mnras] {10.1111/j.1365-2966.2005.09902.x}, \href
  {https://ui.adsabs.harvard.edu/abs/2006MNRAS.366..787K} {366, 787}

\bibitem[\protect\citeauthoryear{{Krajnovi{\'c}} et~al.,}{{Krajnovi{\'c}}
  et~al.}{2008}]{2008MNRAS.390...93K}
{Krajnovi{\'c}} D.,  et~al., 2008, \mn@doi [\mnras]
  {10.1111/j.1365-2966.2008.13712.x}, \href
  {https://ui.adsabs.harvard.edu/abs/2008MNRAS.390...93K} {390, 93}

\bibitem[\protect\citeauthoryear{{Krajnovi{\'c}} et~al.,}{{Krajnovi{\'c}}
  et~al.}{2011}]{2011MNRAS.414.2923K}
{Krajnovi{\'c}} D.,  et~al., 2011, \mn@doi [\mnras]
  {10.1111/j.1365-2966.2011.18560.x}, \href
  {https://ui.adsabs.harvard.edu/abs/2011MNRAS.414.2923K} {414, 2923}

\bibitem[\protect\citeauthoryear{{Krajnovi{\'c}} et~al.,}{{Krajnovi{\'c}}
  et~al.}{2013}]{2013MNRAS.432.1768K}
{Krajnovi{\'c}} D.,  et~al., 2013, \mn@doi [\mnras] {10.1093/mnras/sts315},
  \href {https://ui.adsabs.harvard.edu/abs/2013MNRAS.432.1768K} {432, 1768}

\bibitem[\protect\citeauthoryear{{Lablanche} et~al.,}{{Lablanche}
  et~al.}{2012}]{2012MNRAS.424.1495L}
{Lablanche} P.-Y.,  et~al., 2012, \mn@doi [\mnras]
  {10.1111/j.1365-2966.2012.21343.x}, \href
  {https://ui.adsabs.harvard.edu/abs/2012MNRAS.424.1495L} {424, 1495}

\bibitem[\protect\citeauthoryear{{Leung} et~al.,}{{Leung}
  et~al.}{2018}]{2018MNRAS.477..254L}
{Leung} G. Y.~C.,  et~al., 2018, \mn@doi [\mnras] {10.1093/mnras/sty288}, \href
  {https://ui.adsabs.harvard.edu/abs/2018MNRAS.477..254L} {477, 254}

\bibitem[\protect\citeauthoryear{{Li}, {Li}, {Mao}, {Xu}, {Long}  \&
  {Emsellem}}{{Li} et~al.}{2016}]{2016MNRAS.455.3680L}
{Li} H.,  {Li} R.,  {Mao} S.,  {Xu} D.,  {Long} R.~J.,   {Emsellem} E.,  2016,
  \mn@doi [\mnras] {10.1093/mnras/stv2565}, \href
  {https://ui.adsabs.harvard.edu/abs/2016MNRAS.455.3680L} {455, 3680}

\bibitem[\protect\citeauthoryear{{Li} et~al.,}{{Li}
  et~al.}{2017}]{2017ApJ...838...77L}
{Li} H.,  et~al., 2017, \mn@doi [\apj] {10.3847/1538-4357/aa662a}, \href
  {https://ui.adsabs.harvard.edu/abs/2017ApJ...838...77L} {838, 77}

\bibitem[\protect\citeauthoryear{{Li} et~al.,}{{Li}
  et~al.}{2019}]{2019MNRAS.490.2124L}
{Li} R.,  et~al., 2019, \mn@doi [\mnras] {10.1093/mnras/stz2565}, \href
  {https://ui.adsabs.harvard.edu/abs/2019MNRAS.490.2124L} {490, 2124}

\bibitem[\protect\citeauthoryear{{Lotz}, {Primack}  \& {Madau}}{{Lotz}
  et~al.}{2004}]{2004AJ....128..163L}
{Lotz} J.~M.,  {Primack} J.,   {Madau} P.,  2004, \mn@doi [\aj]
  {10.1086/421849}, \href
  {https://ui.adsabs.harvard.edu/abs/2004AJ....128..163L} {128, 163}

\bibitem[\protect\citeauthoryear{{Lovell} et~al.,}{{Lovell}
  et~al.}{2018}]{2018MNRAS.481.1950L}
{Lovell} M.~R.,  et~al., 2018, \mn@doi [\mnras] {10.1093/mnras/sty2339}, \href
  {https://ui.adsabs.harvard.edu/abs/2018MNRAS.481.1950L} {481, 1950}

\bibitem[\protect\citeauthoryear{{Naab} et~al.,}{{Naab}
  et~al.}{2014}]{2014MNRAS.444.3357N}
{Naab} T.,  et~al., 2014, \mn@doi [\mnras] {10.1093/mnras/stt1919}, \href
  {https://ui.adsabs.harvard.edu/abs/2014MNRAS.444.3357N} {444, 3357}

\bibitem[\protect\citeauthoryear{{Nelson} et~al.,}{{Nelson}
  et~al.}{2019a}]{2019ComAC...6....2N}
{Nelson} D.,  et~al., 2019a, \mn@doi [Computational Astrophysics and Cosmology]
  {10.1186/s40668-019-0028-x}, \href
  {https://ui.adsabs.harvard.edu/abs/2019ComAC...6....2N} {6, 2}

\bibitem[\protect\citeauthoryear{{Nelson} et~al.,}{{Nelson}
  et~al.}{2019b}]{2019MNRAS.490.3234N}
{Nelson} D.,  et~al., 2019b, \mn@doi [\mnras] {10.1093/mnras/stz2306}, \href
  {https://ui.adsabs.harvard.edu/abs/2019MNRAS.490.3234N} {490, 3234}

\bibitem[\protect\citeauthoryear{{Peng}, {Ho}, {Impey}  \& {Rix}}{{Peng}
  et~al.}{2010}]{2010AJ....139.2097P}
{Peng} C.~Y.,  {Ho} L.~C.,  {Impey} C.~D.,   {Rix} H.-W.,  2010, \mn@doi [\aj]
  {10.1088/0004-6256/139/6/2097}, \href
  {https://ui.adsabs.harvard.edu/abs/2010AJ....139.2097P} {139, 2097}

\bibitem[\protect\citeauthoryear{{Pillepich} et~al.,}{{Pillepich}
  et~al.}{2019}]{2019MNRAS.490.3196P}
{Pillepich} A.,  et~al., 2019, \mn@doi [\mnras] {10.1093/mnras/stz2338}, \href
  {https://ui.adsabs.harvard.edu/abs/2019MNRAS.490.3196P} {490, 3196}

\bibitem[\protect\citeauthoryear{{Rix} \& {Zaritsky}}{{Rix} \&
  {Zaritsky}}{1995}]{1995ApJ...447...82R}
{Rix} H.-W.,  {Zaritsky} D.,  1995, \mn@doi [\apj] {10.1086/175858}, \href
  {https://ui.adsabs.harvard.edu/abs/1995ApJ...447...82R} {447, 82}

\bibitem[\protect\citeauthoryear{{Rousseeuw} \& {Van Driessen}}{{Rousseeuw} \&
  {Van Driessen}}{2006}]{ROUSSEEUW2006}
{Rousseeuw} P.~J.,  {Van Driessen} K.,  2006, \mn@doi [Data Mining and
  Knowledge Discovery] {10.1007/s10618-005-0024-4}, 12, 29

\bibitem[\protect\citeauthoryear{{Rudnick} \& {Rix}}{{Rudnick} \&
  {Rix}}{1998}]{1998AJ....116.1163R}
{Rudnick} G.,  {Rix} H.-W.,  1998, \mn@doi [\aj] {10.1086/300518}, \href
  {https://ui.adsabs.harvard.edu/abs/1998AJ....116.1163R} {116, 1163}

\bibitem[\protect\citeauthoryear{{Schwarzschild}}{{Schwarzschild}}{1979}]{1979ApJ...232..236S}
{Schwarzschild} M.,  1979, \mn@doi [\apj] {10.1086/157282}, \href
  {https://ui.adsabs.harvard.edu/abs/1979ApJ...232..236S} {232, 236}

\bibitem[\protect\citeauthoryear{{Scott} et~al.,}{{Scott}
  et~al.}{2018}]{2018MNRAS.481.2299S}
{Scott} N.,  et~al., 2018, \mn@doi [\mnras] {10.1093/mnras/sty2355}, \href
  {https://ui.adsabs.harvard.edu/abs/2018MNRAS.481.2299S} {481, 2299}

\bibitem[\protect\citeauthoryear{{S{\'e}rsic}}{{S{\'e}rsic}}{1963}]{1963BAAA....6...41S}
{S{\'e}rsic} J.~L.,  1963, Boletin de la Asociacion Argentina de Astronomia La
  Plata Argentina, \href
  {https://ui.adsabs.harvard.edu/abs/1963BAAA....6...41S} {6, 41}

\bibitem[\protect\citeauthoryear{{Taylor} et~al.,}{{Taylor}
  et~al.}{2011}]{2011MNRAS.418.1587T}
{Taylor} E.~N.,  et~al., 2011, \mn@doi [\mnras]
  {10.1111/j.1365-2966.2011.19536.x}, \href
  {https://ui.adsabs.harvard.edu/abs/2011MNRAS.418.1587T} {418, 1587}

\bibitem[\protect\citeauthoryear{{Watkins}, {van de Ven}, {den Brok}  \& {van
  den Bosch}}{{Watkins} et~al.}{2013}]{2013MNRAS.436.2598W}
{Watkins} L.~L.,  {van de Ven} G.,  {den Brok} M.,   {van den Bosch} R. C.~E.,
  2013, \mn@doi [\mnras] {10.1093/mnras/stt1756}, \href
  {https://ui.adsabs.harvard.edu/abs/2013MNRAS.436.2598W} {436, 2598}

\bibitem[\protect\citeauthoryear{{Wolf}, {Martinez}, {Bullock}, {Kaplinghat},
  {Geha}, {Mu{\~n}oz}, {Simon}  \& {Avedo}}{{Wolf}
  et~al.}{2010}]{2010MNRAS.406.1220W}
{Wolf} J.,  {Martinez} G.~D.,  {Bullock} J.~S.,  {Kaplinghat} M.,  {Geha} M.,
  {Mu{\~n}oz} R.~R.,  {Simon} J.~D.,   {Avedo} F.~F.,  2010, \mn@doi [\mnras]
  {10.1111/j.1365-2966.2010.16753.x}, \href
  {https://ui.adsabs.harvard.edu/abs/2010MNRAS.406.1220W} {406, 1220}

\bibitem[\protect\citeauthoryear{Zhong}{Zhong}{2023}]{zhzh2023}
Zhong Z.,  2023, {AsymParaEta: Calculating the asymmetric parameter eta},
  \mn@doi{10.5281/zenodo.8270377}

\bibitem[\protect\citeauthoryear{{Zhong}, {Zhao}, {Rix}  \& {Ho}}{{Zhong}
  et~al.}{2023}]{2023ApJ...957L..12Z}
{Zhong} Z.-H.,  {Zhao} G.,  {Rix} H.-W.,   {Ho} L.~C.,  2023, \mn@doi [\apjl]
  {10.3847/2041-8213/acffca}, \href
  {https://ui.adsabs.harvard.edu/abs/2023ApJ...957L..12Z} {957, L12}

\bibitem[\protect\citeauthoryear{{Zhu}, {van de Ven}, {Watkins}  \&
  {Posti}}{{Zhu} et~al.}{2016}]{2016MNRAS.463.1117Z}
{Zhu} L.,  {van de Ven} G.,  {Watkins} L.~L.,   {Posti} L.,  2016, \mn@doi
  [\mnras] {10.1093/mnras/stw2081}, \href
  {https://ui.adsabs.harvard.edu/abs/2016MNRAS.463.1117Z} {463, 1117}

\bibitem[\protect\citeauthoryear{{Zhu}, {Lu}, {Cappellari}, {Li}, {Mao}  \&
  {Gao}}{{Zhu} et~al.}{2023}]{2023MNRAS.522.6326Z}
{Zhu} K.,  {Lu} S.,  {Cappellari} M.,  {Li} R.,  {Mao} S.,   {Gao} L.,  2023,
  \mn@doi [\mnras] {10.1093/mnras/stad1299}, \href
  {https://ui.adsabs.harvard.edu/abs/2023MNRAS.522.6326Z} {522, 6326}

\bibitem[\protect\citeauthoryear{{de Graaff}, {Trayford}, {Franx}, {Schaller},
  {Schaye}  \& {van der Wel}}{{de Graaff} et~al.}{2022}]{2022MNRAS.511.2544D}
{de Graaff} A.,  {Trayford} J.,  {Franx} M.,  {Schaller} M.,  {Schaye} J.,
  {van der Wel} A.,  2022, \mn@doi [\mnras] {10.1093/mnras/stab3510}, \href
  {https://ui.adsabs.harvard.edu/abs/2022MNRAS.511.2544D} {511, 2544}

\bibitem[\protect\citeauthoryear{{de Graaff}, {Franx}, {Bell}, {Bezanson},
  {Schaller}, {Schaye}  \& {van der Wel}}{{de Graaff}
  et~al.}{2023}]{2023MNRAS.518.5376D}
{de Graaff} A.,  {Franx} M.,  {Bell} E.~F.,  {Bezanson} R.,  {Schaller} M.,
  {Schaye} J.,   {van der Wel} A.,  2023, \mn@doi [\mnras]
  {10.1093/mnras/stac3277}, \href
  {https://ui.adsabs.harvard.edu/abs/2023MNRAS.518.5376D} {518, 5376}

\bibitem[\protect\citeauthoryear{{van de Sande} et~al.,}{{van de Sande}
  et~al.}{2017}]{2017ApJ...835..104V}
{van de Sande} J.,  et~al., 2017, \mn@doi [\apj] {10.3847/1538-4357/835/1/104},
  \href {https://ui.adsabs.harvard.edu/abs/2017ApJ...835..104V} {835, 104}

\bibitem[\protect\citeauthoryear{{van der Wel} et~al.,}{{van der Wel}
  et~al.}{2022}]{2022ApJ...936....9V}
{van der Wel} A.,  et~al., 2022, \mn@doi [\apj] {10.3847/1538-4357/ac83c5},
  \href {https://ui.adsabs.harvard.edu/abs/2022ApJ...936....9V} {936, 9}

\makeatother
\end{thebibliography}

\bsp	
\label{lastpage}
\end{document}